\documentstyle[12pt,cite,epsfig]{article}

\renewcommand{\today}{March 15, 1995}

\newcommand{\nc}{\newcommand}
\nc{\be}{\begin{equation}}
\nc{\ee}{\end{equation}}
\nc{\bea}{\begin{eqnarray}}
\nc{\eea}{\end{eqnarray}}
\nc{\beas}{\begin{eqnarray*}}
\nc{\eeas}{\end{eqnarray*}}
\nc{\noi}{\noindent}
\nc{\sD}{\not \! \! D}
\nc{\s}[1]{\not \! #1}
\nc{\non}{\nonumber}
\nc{\bb}{\bibitem}
\nc{\rw}{$\rho\!-\!\omega$ }
\nc{\lf}{\left}
\nc{\r}{\right}
\nc{\mb}[1]{\makebox[#1]{}}
\nc{\pa}{\partial}
\nc{\sA}{\not \! \! A}
\nc{\newsec}[1]{\section{#1}\mb{0.5cm}}
\nc{\h}{\frac{1}{2}}
\nc{\ra}{\rightarrow}
\nc{\la}{\leftarrow}
\nc{\ep}{$e^+e^-\ra\pi^+\pi^-$}
\def\mathunderaccent#1{\let\theaccent#1\mathpalette\putaccentunder}
\def\putaccentunder#1#2{\oalign{$#1#2$\crcr\hidewidth
\vbox to.2ex{\hbox{$#1\theaccent{}$}\vss}\hidewidth}}

\nc{\ti}{\mathunderaccent\tilde}
\nc{\M}{{\cal M}}

\begin{document}
\thispagestyle{empty}
\begin{flushright}
ADP-95-15/T176 \\
hep-ph/9503332
\end{flushright}

\begin{center}
{\large{\bf Rho-omega Mixing and the Pion Electromagnetic Form-Factor}} \\
\vspace{1cm}
{\large{\bf [Phys. Lett. B354 (1995) 14]}}\\
\vspace{2.2 cm}
H.B.~O'Connell, B.C.~Pearce, \\
A.W.~Thomas and A.G.~Williams \\
\vspace{1.2 cm}
{\it
Department of Physics and Mathematical Physics \\
University of Adelaide, S.Aust 5005, Australia } \\
\vspace{1.2 cm}
\today
\vspace{1.2 cm}
\begin{abstract}

The suggestion of momentum dependence in the amplitude for rho-omega mixing has
generated concern over related implications for vector meson dominance and the
photon-rho coupling.  We discuss two established representations of vector
meson dominance and show that one of these is completely consistent with such a
coupling. We then apply it to a calculation of the  pion electromagnetic
form-factor.  Our analysis leads to a new value for the on-shell rho-omega
mixing amplitude of $-3800\pm 370\;{\rm MeV}^2$.

\end{abstract}

\end{center}
\vspace{2.5cm}
\begin{flushleft}
E-mail: {\it hoconnel, bpearce, athomas, awilliam@physics.adelaide.edu.au}

\end{flushleft}

\newpage

\section{Introduction}
\mb{.5cm}

It was recently suggested that the \rw mixing amplitude, $\Pi_{\rho\omega}$,
might vary fairly rapidly with the invariant mass-squared of the meson near
$q^2=0$ \cite{GHT}. Since then there has been  a significant amount of work on
the problem,  \cite{KTW,MTRC,PW,HHMK,Re}, culminating in the suggestion that
under certain reasonable conditions $\Pi_{\rho\omega}$ should actually vanish
at $q^2=0$ \cite{OPTW}.  As we will show explicitly, this behaviour
presents no difficulty with regard to our understanding of the mixing amplitude
at the $\omega$ pole, through the interference of the decay $\omega\ra\pi\pi$
with $\rho\ra\pi\pi$ \cite{CSM,us}. However, it has profound implications for
the modelling of charge symmetry violation (CSV) in the nucleon--nucleon ($NN$)
force \cite{CB,MSC,MilWil}.  Indeed the  variation found in Refs.
\cite{GHT,KTW,MTRC,PW,HHMK,Re,OPTW} dramatically reduces the conventional CSV
$NN$ potentials obtained under the usual assumption that $\Pi_{\rho\omega}$ is
a constant \cite{GHT,IN}.

Although there is no rigorous derivation of the coupling of the photon to the
$\rho$ in QCD, if it occurs through the same sort of quark loop used in
Refs.~\cite{GHT,KTW,MTRC,PW} to model \rw mixing one would expect that it too
should vanish at $q^2=0$. This is in contrast with the most commonly used
version of vector meson dominance (VMD) where the $\gamma-\rho$ coupling is
fixed (see below). As discussed by Miller \cite{MO}, it is not obvious that one
can reconcile a $\gamma-\rho$ mixing amplitude that vanishes at $q^2=0$ with
the success of VMD in describing data such as the pion form-factor measured in
\ep. Clearly the resolution of this issue is essential to our understanding of
CSV.

Our purpose here is to show that one can fit the measured pion form-factor with
a $\gamma-\rho$ coupling that vanishes at $q^2=0$. We begin by recalling that
there are actually two related representations of VMD. We shall see that the
one less frequently used, described by Sakurai in the 1960's, is indeed
consistent with this vanishing coupling and reproduces the form-factor very
nicely. In the process of fitting the data, we extract a new value of the \rw
mixing amplitude at the $\omega$ pole, $\Pi_{\rho\omega}(m_\omega^2)$.

\section{Vector Meson Dominance}
\mb{.5cm}

To examine \rw mixing in any detail requires an understanding of the model used
to describe the process. Our available data on \rw mixing is almost exclusively
obtained from the electromagnetic (EM) pion form-factor.  Like almost all low
energy photon-hadron interactions \cite{W} the EM form-factor is modelled by
VMD \cite{Sak2}, which we shall now discuss.

VMD assumes that the dominant role in the interaction of the photon with
hadronic matter is played by vector mesons. It is an attempt to model
non-perturbative interactions determined by QCD, which,  cannot yet be
evaluated in this low-energy regime. The traditional representation of VMD
(which, following the conventions of our recent review \cite{us}, we shall
refer to as VMD2) assumes that the photon couples to hadronic matter {\em
exclusively} through a vector meson, to which it couples with a {\em fixed}
strength proportional to the mass squared of the meson.

For the photon--rho--pion system, the relevant part of the VMD2 Lagrangian
is
\be
{\cal L}_{\rm VMD2}= -\frac{1}{4}F_{\mu\nu}F^{\mu\nu}-\frac{1}{4}
\rho_{\mu\nu}\rho^{\mu\nu}+\h m_\rho^2(\rho_\mu)^2 -g_{\rho\pi\pi}\rho_\mu
J^\mu_\pi-\frac{e m_\rho^2}{g_\rho}\rho_\mu
A^\mu+\h\left(\frac{e}{g_\rho}\right)^2 m_\rho^2 A_\mu A^\mu,
\label{newvmd}
\ee
where $J^\mu_\pi$ is the pion current, $(\vec{\pi}\times\pa_\mu\vec{\pi})_3$,
and $F_{\mu\nu}$ and $\rho_{\mu\nu}$ are the EM and $\rho$ field strength
tensors (here $e\equiv|e|$). {From} Eq.~(\ref{newvmd}) one arrives at a pion
form-factor of the form \cite{us}
\be
F_\pi(q^2)=-\frac{m_\rho^2}{q^2-m_\rho^2+im_\rho\Gamma_\rho(q^2)}
\frac{g_{\rho\pi\pi}}{g_\rho}, \label{ff2}
\ee
where conventionally one takes \cite{Bark,DSM}
\be
\Gamma_\rho(q^2)=\Gamma_\rho
\left(\frac{q^2-4m_\pi^2}{m_\rho^2-4m_\pi^2}\right)^{3/2}
\frac{m_\rho}{\sqrt{q^2}}\theta(q^2-4m_\pi^2).
\ee
This VMD2 Lagrangian, rederived by Bando {\it et al.} \cite{Ban} from a model
based on hidden local gauge symmetry, has some unappealing features. Firstly,
the $\rho-\gamma$ interaction is supposed to be modelling the
quark-polarisation of the photon, which necessarily vanishes at $q^2=0$ to
preserve EM gauge invariance \cite{Ro}, whereas the coupling determined by
Eq.~(\ref{newvmd}) is fixed. Hence the VMD2 dressing of the photon propagator
shifts the pole away from zero, and thus a bare photon mass must be
introduced into the Lagrangian to counterbalance this and ensure that the
dressed photon is massless. Secondly, recent studies \cite{Ben,BCP} have shown
that the best fit to \ep requires a non-resonant term (i.e., a contribution in
which the $\rho$ does not appear), which VMD2 lacks. Thirdly, the constraint
\be
F_\pi(0)=1,
\label{zero}
\ee
which reflects the fact that the photon  sees only the charge of the pion at
zero momentum
transfer, is only realised by  Eq.~(\ref{ff2}) in the
limit of universality ($g_\rho=g_{\rho\pi\pi}$), which is seen to be only
approximate in nature \cite{HS}.

For these reasons we prefer the alternative formulation \cite{Sak2} which we
shall call VMD1 \cite{us}, with the following Lagrangian
\be
{\cal L}_{\rm VMD1}=-\frac{1}{4}F_{\mu\nu}F^{\mu\nu}-\frac{1}{4}
\rho_{\mu\nu}\rho^{\mu\nu}+\h m_\rho^2\rho_\mu\rho^\mu- g_{\rho\pi\pi}\rho_\mu
J_\pi^\mu -eA_\mu J_\pi^\mu-\frac{e}{2g_\rho} F_{\mu\nu}\rho^{\mu\nu}.
\label{vmdlag}
\ee
The key features of this representation are the absence of a photon mass term
and the presence of a term $F_{\mu\nu}\rho^{\mu\nu}$, which produces a
momentum-dependent $\gamma-\rho$ coupling of the form \cite{us},
\bea
\nonumber
{\cal L}_{\gamma\rho}&=&-\frac{e}{2g_\rho} F_{\mu\nu}\rho^{\mu\nu} \\
&\ra&-\frac{e}{g_\rho}q^2A_\mu\rho^\mu.
\eea
This, of course, decouples the photon from the $\rho$ at $q^2=0$, hence keeping
the photon  massless in a natural way. However, the photon is still able to
couple to
the hadronic current through the direct coupling $-eA_\mu J^\mu_\pi$, giving us
a non-resonant term. We now have a form-factor of the form \cite{us}
\be
F_\pi(q^2)=1-
\frac{q^2 g_{\rho\pi\pi}}{g_\rho[q^2-m_\rho^2+im_\rho\Gamma_\rho(q^2)]}.
\label{ff1}
\ee
We note that Eq.~(\ref{ff1}) automatically satisfies the requirement for the
form-factor given in Eq.~(\ref{zero}).

\begin{figure}[htb]
  \centering{\
     \epsfig{angle=270,figure=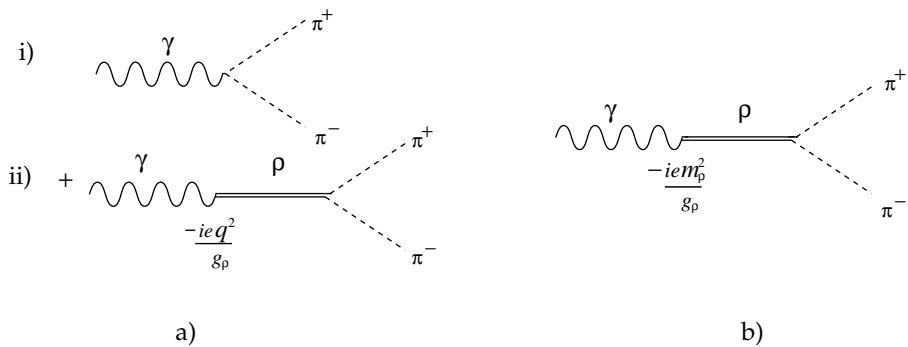,height=5.5cm}  }
\parbox{130mm}{\caption{Contributions to the
pion form-factor in the two representations of vector meson dominance
a) VMD1 b) VMD2.}
\label{twoways}}
\end{figure}

We illustrate the difference between the
two representations in Fig.~\ref{twoways}.

\section{\rw mixing}
\mb{.5cm}
At present the widely quoted value of for $\Pi_{\rho\omega}\equiv
\Pi_{\rho\omega}(m_\omega^2)$ \cite{CB},
is obtained from the branching ratio formula for the $\omega$
\be
B(\omega\ra\pi\pi)=\Gamma(\omega\ra\pi\pi)/\Gamma(\omega),
\label{coon}
\ee
derived from a \rw mixing analysis where
\be
\Gamma(\omega\ra\pi\pi)=
\left|\frac{\Pi_{\rho\omega}}{im_\rho\Gamma_\rho}\right|^2\Gamma
(\rho\ra\pi\pi).
\label{coon2}
\ee
Using the branching ratio determined in 1985 by the
Novosibirsk group \cite{Bark}
\be
B(\omega\ra\pi\pi)=(2.3\pm 0.4\pm 0.2)\%
\ee
Coon and Barrett obtained $\Pi_{\rho\omega}=-4520\pm 600 {\rm MeV}^2$.
We aim to extract
$\Pi_{\rho\omega}$ from a fit to the cross-section of the reaction \ep using
\be
\sigma(q^2)=\frac{\alpha^2\pi}{3}\frac{(q^2-4m_\pi^2)^{3/2}}{s^{5/2}}
|F_\pi(q^2)|^2,
\ee
and the form-factor determined by VMD1 (Eq.~(\ref{ff1})).

So far, we have not introduced any effects of charge symmetry violation (CSV)
into our system, and hence the $\omega$ (which cannot otherwise couple to a
$\pi^+\pi^-$ state) does not appear. In their recent examination of the EM pion
and nucleon form-factors using VMD1, D\"{o}nges {\it et al.} \cite{DSM}
introduced the $\omega$ through a covariant derivative in the pion kinetic
term. This produces a direct contribution from $\omega\ra 2\pi$ without any \rw
mixing, but does not provide a good representation of data in the resonance
region. We shall use the mixed propagator \cite{OPTW}, where the mixing is
introduced by an off-diagonal piece, $\Pi_{\rho\omega}$ in the vector meson
self-energy. To first order in CSV [i.e., to ${\cal O}(\Pi_{\rho\omega}$)], the
propagator is given by (we ignore pieces proportional to $q_\mu$ as we couple
to conserved currents)
\be
D_{\mu\nu}=\left( \begin{array}{cc}
          1/s_{\rho} & \Pi_{\rho\omega}/s_{\rho}s_{\omega} \\
          \Pi_{\rho\omega}/s_{\rho}s_{\omega} & 1/s_{\omega}
         \end{array} \right)\;g_{\mu\nu},
\label{prop3}
\ee
where
\bea
s_\rho&\equiv&q^2-\Pi_{\rho\rho}(q^2)-m_\rho^2, \\
&\equiv&q^2-m_\rho^2+im_\rho\Gamma_\rho(q^2),
\eea
and similarly for the $\omega$. We can now examine how to use this propagator
to model a system with CSV \cite{us}. Note that as the pion form-factor is only
sensitive to \rw mixing near the $\omega$ pole we can ignore the momentum
dependence of the meson mixing and treat $\Pi_{\rho\omega}$ as a constant.

\begin{figure}[htb]
  \centering{\
     \epsfig{angle=270,figure=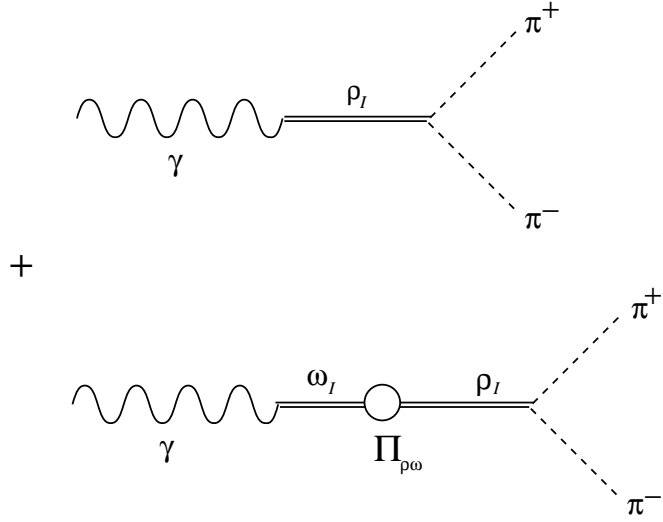,height=7cm}  }
\parbox{130mm}{\caption{The contribution of \rw mixing to the pion form-factor
from the states of pure isospin.}
\label{tony}}
\end{figure}

In a matrix notation, the Feynman amplitude for the process
$\gamma\ra\pi\pi$, proceeding via vector mesons, can be written in
the form
\begin{equation}
i\M_{\mu}^{\gamma\ra\pi\pi} =
  \left( \begin{array}{cc}
          i\M^{\nu}_{\rho_I\ra\pi\pi} & i\M^{\nu}_{\omega_I\ra\pi\pi}
         \end{array} \right)
  iD_{\nu\mu}
  \left( \begin{array}{c}
          i\M_{\gamma\ra\rho_I} \\
          i\M_{\gamma\ra\omega_I}
         \end{array} \right),
\label{Fpimatrix}
\end{equation}
where the matrix $D_{\nu\mu}$ is given by Eq.~(\ref{prop3}) and the other
Feynman amplitudes are derived from ${\cal L}_{\rm VMD1}$.
 If we make the standard assumption that
the pure isospin state $\omega_I$ does not couple to two pions
($\M^{\nu}_{\omega_I\ra\pi\pi}=0$) then to lowest order in the mixing,
Eq.~(\ref{Fpimatrix}) is just
\begin{equation}
\M^{\mu}_{\gamma\ra\pi\pi} =
  \left( \begin{array}{cc}
          \M^{\mu}_{\rho_I\ra\pi\pi} & 0
         \end{array} \right)
  \left( \begin{array}{cc}
          1/s_{\rho} & \Pi_{\rho\omega}/s_{\rho}s_{\omega} \\
          \Pi_{\rho\omega}/s_{\rho}s_{\omega} & 1/s_{\omega}
         \end{array} \right)
  \left( \begin{array}{c}
          \M_{\gamma\ra\rho_I} \\
          \M_{\gamma\ra\omega_I}
         \end{array} \right).
\label{ffampl}
\end{equation}
Expanding this just gives
\begin{equation}
  \M^{\mu}_{\gamma\ra\pi\pi} =
    \M^{\mu}_{\rho_I\ra\pi\pi} \frac{1}{s_{\rho}} \M_{\gamma\ra\rho_I}
    + \M^{\mu}_{\rho_I\ra\pi\pi} \frac{1}{s_{\rho}} \Pi_{\rho\omega}
        \frac{1}{s_{\omega}} \M_{\gamma\ra\omega_I},
\label{nondiag}
\end{equation}
which we recognise as the sum of the two diagrams shown in Fig.~\ref{tony}.

The couplings that enter this expression, through
$\M^{\mu}_{\rho_I\ra\pi\pi}$, $\M_{\gamma\ra\rho_I}$ and
$\M_{\gamma\ra\omega_I}$, always involve the unphysical pure isospin states
$\rho_I$ and $\omega_I$.  However, we can re-express Eq.~(\ref{nondiag}) in
terms of the physical states by first diagonalising the vector meson
propagator. We can, at this point,
introduce a diagonalising matrix \cite{us}
\begin{equation}
C = \left(\begin{array}{cc} 1 & \epsilon \\
-\epsilon & 1 \end{array} \right)
\end{equation}
where, to lowest order in the mixing,
\begin{equation}
\epsilon=\frac{\Pi_{\rho\omega}}{s_\rho-s_\omega}.
\label{epseq}
\end{equation}
We now insert identities into Eq.~(\ref{ffampl}) and obtain
\begin{eqnarray}
\nonumber
\M^{\mu}_{\gamma\ra\pi\pi} &=&
  \left( \begin{array}{cc}
          \M^{\mu}_{\rho_I\ra\pi\pi} & 0
         \end{array} \right)
  C C^{-1}
  \left( \begin{array}{cc}
          1/s_{\rho} & \Pi_{\rho\omega}/s_{\rho}s_{\omega} \\
          \Pi_{\rho\omega}/s_{\rho}s_{\omega} & 1/s_{\omega}
         \end{array} \right)
  C C^{-1}
  \left( \begin{array}{c}
          \M_{\gamma\ra\rho_I} \\
          \M_{\gamma\ra\omega_I}
         \end{array} \right), \\
 &=&
  \left( \begin{array}{cc}
          \M^{\mu}_{\rho\ra\pi\pi} & \M^{\mu}_{\omega\ra\pi\pi}
         \end{array} \right)
  \left( \begin{array}{cc}
          1/s_{\rho} & 0 \\
          0 & 1/s_{\omega}
         \end{array} \right)
  \left( \begin{array}{c}
          \M_{\gamma\ra\rho} \\
          \M_{\gamma\ra\omega}
         \end{array} \right),
\label{diagff}
\end{eqnarray}
where to first order in $\epsilon$ we have identified the physical amplitudes
as
\begin{eqnarray}
\M^{\mu}_{\rho\ra\pi\pi}&=&\M^{\mu}_{\rho_I\ra\pi\pi}, \\
\M^{\mu}_{\omega\ra\pi\pi}&=&\epsilon \M^{\mu}_{\rho_I\ra\pi\pi}, \\
\M_{\gamma\ra\rho}&=&\M_{\gamma\ra\rho_I} - \epsilon \M_{\gamma\ra\omega_I}, \\
\M_{\gamma\ra\omega}&=&\M_{\gamma\ra\omega_I} + \epsilon \M_{\gamma\ra\rho_I}.
\end{eqnarray}
Expanding Eq.~(\ref{diagff}), we find
\begin{eqnarray}
\nonumber
  \M^{\mu}_{\gamma\ra\pi\pi} &=&
    \M^{\mu}_{\rho\ra\pi\pi} \frac{1}{s_{\rho}} \M_{\gamma\ra\rho}
   + \M^{\mu}_{\omega\ra\pi\pi} \frac{1}{s_{\omega}} \M_{\gamma\ra\omega} \\
 &=&
  \M^{\mu}_{\rho\ra\pi\pi} \frac{1}{s_{\rho}} \M_{\gamma\ra\rho}
   + \M^{\mu}_{\rho\ra\pi\pi} \frac{\Pi_{\rho\omega}}{s_\rho-s_\omega}
        \frac{1}{s_{\omega}} \M_{\gamma\ra\omega} ,
\label{diagff2}
\end{eqnarray}
which is the form usually seen in older works. At first glance there seems to
be a slight discrepancy between Eqs.~(\ref{nondiag}) and (\ref{diagff2}). The
source of this is the definition used for the coupling of the vector meson to
the photon. The first, Eq.~(\ref{nondiag}), uses couplings to pure isospin
states, the second, Eq.~(\ref{diagff2}) uses ``physical" couplings (i.e.,
couplings to the mass eigenstates) which introduce a leptonic contribution to
the Orsay phase, as discussed by Coon {\it et al.}  \cite{CSM}. This phase is,
however, rather small.  If we assume
$\M_{\gamma\ra\rho_I}=3\M_{\gamma\ra\omega_I}$ and define the leptonic phase
$\theta$ by
\begin{equation}
\frac{\M_{\gamma\ra\omega}}{\M_{\gamma\ra\rho}}=\frac{1}{3}e^{i\theta}
\label{lep1}
\end{equation}
then, to order $\epsilon$,
\begin{equation}
\tan\theta=\frac{10\Pi_{\rho\omega}}{3m_\rho\Gamma_\rho}.
\label{lep2}
\end{equation}
This gives $\theta=5.7^{\rm o}$ for $\Pi_{\rho\omega}=-4520 {\rm MeV}^2$, as
obtained by Coon {\it et al.}  \cite{CSM}.  This small leptonic contribution to
the Orsay phase is the principal manifestation of diagonalising the \rw
propagator.

\section{Fit to $\sigma$(\ep) using VMD1}
\mb{.5cm}

We are now in a position to write down the CSV form-factor based on the VMD1
form-factor of Eq.~(\ref{ff1}) and the mixed state contribution of
Eq.~(\ref{diagff2}),
\be
  F_{\pi}(q^2) = 1 - \frac{q^2 g_{\rho\pi\pi}}{
        g_{\rho} [q^2 - m_\rho^2 + im_\rho \Gamma_\rho(q^2)]}
      - \frac{q^2 \epsilon g_{\rho\pi\pi}}{
        g_{\omega} [q^2 - m_\omega^2 + im_\omega\Gamma_\omega]}
\label{final}
\ee
where,
\bea
  \epsilon &=& \frac{\Pi_{\rho\omega}}{s_\rho-s_\omega} \\
&=&\frac{\Pi_{\rho\omega}}{
        m_\omega^2 - m_\rho^2
        - i(m_\omega\Gamma_\omega - m_\rho\Gamma_\rho(q^2))}.
\eea
The $\omega$ decay formula of Eq.~(\ref{coon2}) can now be seen to follow from
Eq.~(\ref{final}) with an approximation (namely that $\Gamma_\omega$ is very
small and that $m_\rho^2=m_\omega^2$) for $\epsilon$.  Because the width of the
$\omega$ is very small we can safely neglect any  momentum dependence in it,
and simply use $\Gamma_\omega(m_\omega^2)$ \cite{us,Ben}. In principle we
should include a contribution in Eq.~(\ref{final}) from the leptonic phase
(Eqs.~(\ref{lep1}) and (\ref{lep2})), but, as this is very small, we can
also safely ignore it and assume all phase comes from $\epsilon$.

\begin{figure}[htb]
  \centering{\
     \epsfig{angle=0,figure=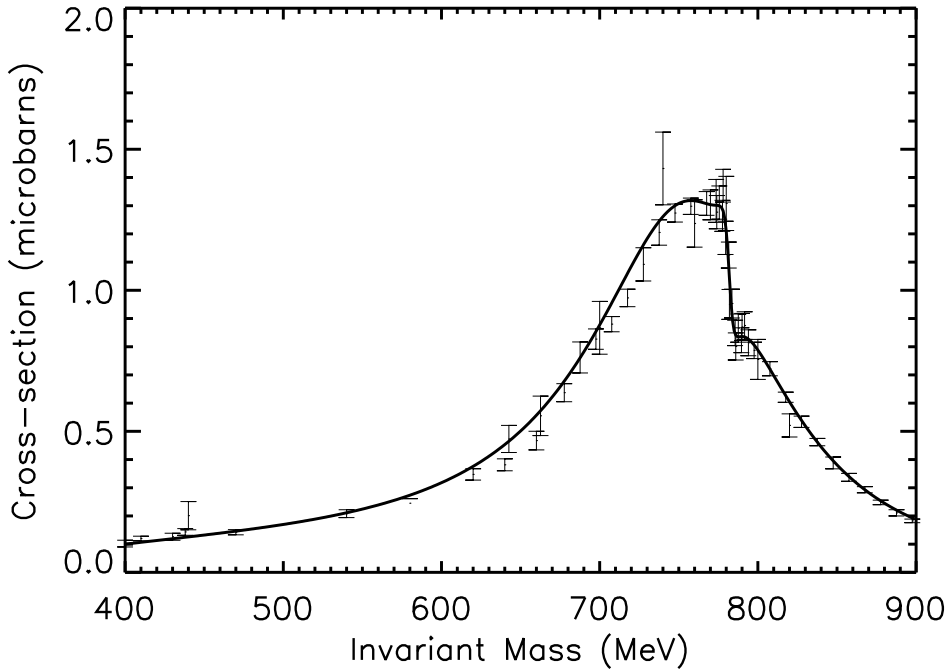,height=10cm}  }
\parbox{130mm}{\caption{Cross-section of $e^+e^-\ra\pi^+\pi^-$ plotted as a
function of the energy in the centre of mass. The experimental data is from
Refs.~[17,21].}
\label{graph1}}
\end{figure}

\begin{figure}[htb]
  \centering{\
     \epsfig{angle=0,figure=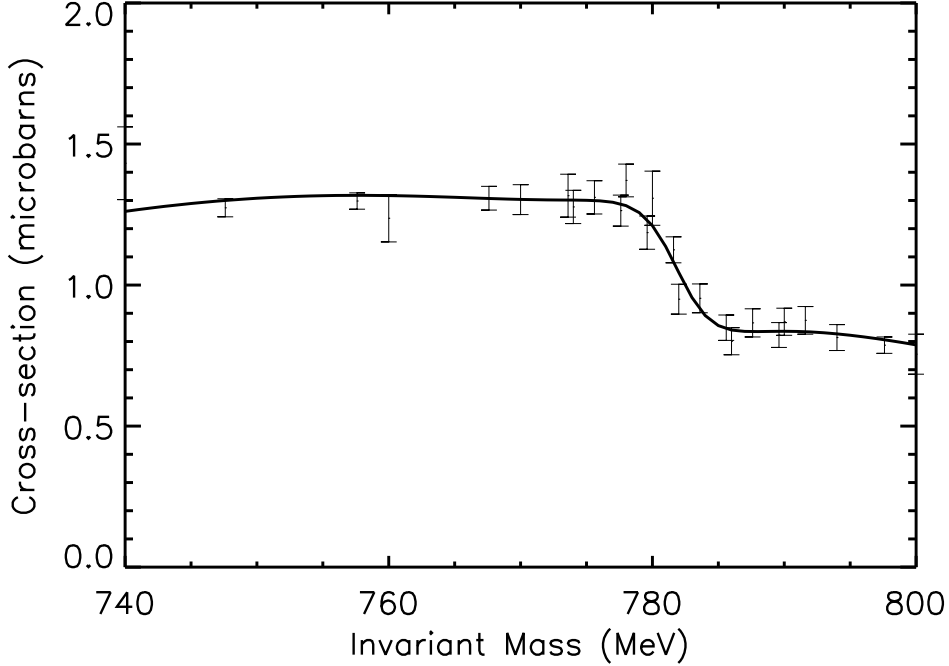,height=10cm}  }
\parbox{130mm}{\caption{Cross-section for $e^+e^-\ra\pi^+\pi^-$ in the region
around the resonance where \rw mixing is most noticeable. The experimental
data is from Refs.~[17,21].}
\label{graph2}}
\end{figure}

All parameters except $\Pi_{\rho\omega}$ are fixed by various data as
discussed below. The results of fitting this remaining parameter to the data
are shown in Fig.~\ref{graph1} with the resonance region shown
in close-up in Fig.~\ref{graph2}.

The mass and width of the $\omega$ are as given by the Particle Data Group
(PDG) \cite{PDG}, $m_\omega=781.94\pm 0.12$ MeV and $\Gamma_\omega=8.43\pm
0.10$ MeV.  There has recently been considerable interest in the value of the
$\rho$ parameters, $m_\rho$ and $\Gamma_\rho$ with studies showing that the
optimal values \cite{Ben,BCP} may differ slightly from those given by the PDG.
The value of $\Pi_{\rho\omega}$ is not sensitive to the masses and widths, and
we have obtained a good fit with $m_\rho=772$MeV and $\Gamma_\rho$=149 MeV,
which are  close to the PDG values.

The values of the coupling constants are however quite important for an
extraction of $\Pi_{\rho\omega}$. We obtain $g_\rho$ and $g_{\rho\pi\pi}$ from
$\Gamma(\rho\ra e^+ e^-)\sim 6.8$ MeV and $\Gamma(\rho\ra\pi\pi)\sim 149$ MeV
\bea
g_{\rho\pi\pi}^2/4\pi&\sim& 2.9, \\
\label{val1}
g_{\rho}^2/4\pi&\sim& 2.0,
\label{val2}
\eea
which show, for example, that universality is not strictly obeyed (as
mentioned previously).

Historically the ratio $g_\omega/g_\rho$ was believed to be around 3
\cite{GFQ}, a figure supported in a recent QCD-based analysis \cite{DM}.
Empirically though, the ratio can be determined \cite{BCP} from leptonic
partial rates \cite{PDG} giving
\bea
\frac{g_\omega}{g_\rho}&=&
\sqrt{\frac{m_\omega\Gamma(\rho\ra e^+e^-)}{m_\rho\Gamma(\omega\ra e^+e^-)}} \\
&=&3.5\pm 0.18.
\label{cuperror}
\eea
Using these parameters we obtain a best fit around the resonance region shown
in Fig.~\ref{graph2} ($\chi^2/{\rm d.o.f.}=14.1/25$) with
$\Pi_{\rho\omega}=-3800$.
There are two principle
sources of error in our value for $\Pi_{\rho\omega}$. The first is a
statistical uncertainty of $310\; {\rm MeV}^2$ for the fit to data, and the
second, of approximately $200\; {\rm MeV}^2$ is due to the error quoted in
Eq.~(\ref{cuperror}). Adding these in quadrature gives us a final value
for the {\em total} mixing amplitude, to be compared with the value $-4520
\pm600\;{\rm MeV}^2$ obtained by Coon and Barrett \cite{CB}. We find
\be
\Pi_{\rho\omega}=-3800\pm 370\; {\rm MeV}^2.
\ee

\section{Conclusions}
\mb{.5cm}
It is now clear that a momentum dependent
$\gamma^*-\rho$ coupling, together with a direct coupling of the photon to
hadronic matter, yields an entirely adequate model of the pion form-factor.
In fact, this picture is basically suggested by attempts to examine the
$\gamma^*-\rho$ coupling via a quark loop. Model calculations typically find
that the loop is momentum-dependent, and vanishes at $q^2=0$ (unless gauge
invariance is spoiled by form-factors, or something of this nature). However,
coupling the photon to quarks in the loop implies that the photon must
also couple to the quarks in hadronic matter, thus introducing a direct
photon-hadron coupling (independent of the $\rho$-meson), and leads us to take
VMD1 as the preferred representation of vector meson dominance. It should now
be clear that the appropriate representation of vector meson dominance to be
used in combination with mixing amplitudes that vanish at $q^2=0$ is VMD1. The
use of VMD2 in this context is inconsistent. As long as one is clear on this
point, there are no dire consequences for momentum dependence in \rw mixing.

\vspace{2.5cm}
{\bf Acknowledgments}
\mb{.5cm}

This work was supported by the Australian Research Council.


\begin{thebibliography}{99}
\bibitem{GHT} T.~Goldman, J.A.~Henderson and A.W.~Thomas, Few Body Systems
{\bf 12},   123 (1992).
\bibitem{KTW} G.~Krein, A.W.~Thomas and A.G.~Williams, Phys. Lett. B {\bf 317},
293 (1993).
\bibitem{MTRC} K.L.~Mitchell, P.C.~Tandy, C.D.~Roberts and R.T.~Cahill,
Phys. Lett. B {\bf 335}, 282 (1994).
\bibitem{PW} J.~Piekarawicz and A.G.~Williams,
Phys. Rev. C {\bf 47}  R2462 (1993).
\bibitem{HHMK} T.~Hatsuda, E.M.~Henley, Th.~Meissner and G.~Krein,
Phys. Rev. C {\bf 49}, 452 (1994).
\bibitem{Re} R.~Friedrich and H.~Reinhardt, \rw mixing and the pion
electromagnetic form-factor in the Nambu--Jona-Lasinio Model, hep-ph/9501333.
\bibitem{OPTW} H.B.~O'Connell, B.C.~Pearce, A.W.~Thomas and A.G.~Williams,
Phys. Lett. B {\bf 336}, 1 (1994).
\bibitem{CSM} S.A.~Coon, M.D.~Scadron and P.C.~Mc~Namee, Nucl. Phys.
{\bf A287}, 381 (1977).
\bb{us} H.B.~O'Connell, B.C.~Pearce, A.W.~Thomas and A.G.~Williams,
hep-ph/9501251, submitted for publication in
{\it Trends in Particle and Nuclear Physics}, ed. W-Y Pauchy Hwang
(Plenum Press).
\bibitem{CB} S.A.~Coon and R.C.~Barrett, Phys. Rev. C {\bf 36}, 2189 (1987).
\bibitem{MSC} P.C.~McNamee, M.D.~Scadron and S.A.~Coon, Nucl. Phys. {\bf A249},
483 (1975).
\bibitem{MilWil} G.A.~Miller, A.W.~Thomas and A.G.~Williams,
Phys. Rev. Lett. {\bf 56}, 2567 (1986); \\
A.G.~Williams, A.W.~Thomas and G.A.~Miller,
Phys. Rev. C {\bf 36}, 1956 (1987).
\bibitem{IN} M.J.~Iqbal and J.A.~Niskanen,
Phys. Lett. B {\bf 322}, 7 (1994).
\bibitem{MO} G.A.~Miller and W.T.H~van~Oers,  nucl-th/9409013, Chapter for
{\em Symmetries and Fundamental Interactions
in Nuclei}, eds. E.M. Henley and W. Haxton (World Scientific).
\bibitem{W} W.~Weise, Phys. Rep. {\bf 13}, 53 (1974).
\bibitem{Sak2} J.J.~Sakurai, {\it Currents and Mesons}, University
of Chicago Press (1969).
\bibitem{Bark} L.M.~Barkov {\it et al.}, Nucl. Phys. {\bf B256}  365 (1985).
\bibitem{DSM} H.C.~D\"{o}nges, M.~Sch\"{a}fer and U.~Mosel, nucl-th/9407012,
to appear in Phys. Rev C, Feb 1995.
\bibitem{Ban} M.~Bando {\it et al.} Phys. Rev. Lett. {\bf 54}, 1215 (1985).
\bibitem{Ro} C.D.~Roberts, Electromagnetic Pion Form Factor and
Neutral Pion Decay Width, hep-ph/9408233.
\bibitem{Ben} D.~Benaksas {\it et al.}, Phys. Lett. {\bf 39B}, 289 (1972).
\bibitem{BCP} A.~Bernicha, G.~L\'{o}pez Castro and J.~Pestieau, Phys. Rev. D
{\bf 50}, 4454 (1994)
\bibitem{HS}  T.~Hakioglu and M.D.~Scadron, Phys. Rev. D
{\bf 43}, 2439 (1991)
\bibitem{PDG} Particle Data Group, Phys. Rev. D {\bf 50}, 1173 (1994).
\bibitem{GFQ} A.S.~Goldhaber, G.C.~Fox and C.~Quigg, Phys. Lett. {\bf 30B},
249 (1969).
\bibitem{DM} G.~Dillon and G.~Morpurgo, Zeit. Phys. C {\bf 46}, 467 (1994).


\end{thebibliography}
\end{document}